\documentclass[prd,twocolumn,superscriptaddress,floatfix,amsmath,footinbib,amssymb]{revtex4-2}
\pdfoutput=1
\usepackage{amssymb}
\usepackage{amsmath}
\usepackage{amsfonts}
\usepackage{tikz}
\usepackage{bm}
\usepackage{physics}
\usepackage{mathtools}
\usepackage{appendix}
\usepackage{adjustbox}
\usepackage{float}

\usepackage{epsfig}
\usepackage{t1enc}
\usepackage{soul}
\usepackage{color}

\usepackage{pgfplots}
\pgfplotsset{compat=1.8}
\usetikzlibrary{arrows.meta}

\usepackage{mathrsfs}
\usepackage{url}
\usepackage[compat=1.1.0]{tikz-feynman}
\usepackage{feynmp-auto}

\usepackage{hyperref}
\hypersetup{
    colorlinks=true,
    linkcolor=green,
    filecolor=magenta,      
    urlcolor=blue,
    breaklinks=true
}

\begin{document}

\title{Higgs-like field interactions before symmetry breaking}

\date{\today}

\author{Jerzy Paczos}
\email{jerzy.paczos@fysik.su.se}
\affiliation{Department of Physics, Stockholm University, SE-106 91 Stockholm, Sweden}

\author{Szymon Cedrowski}
\email{s.cedrowski@student.uw.edu.pl}
\affiliation{Institute of Theoretical Physics, University of Warsaw, Pasteura 5, 02-093 Warsaw, Poland}

\author{Krzysztof Turzyński}
\email{Krzysztof-Jan.Turzynski@fuw.edu.pl}
\affiliation{Institute of Theoretical Physics, University of Warsaw, Pasteura 5, 02-093 Warsaw, Poland}

\author{Andrzej Dragan}
\email{dragan@fuw.edu.pl}
\affiliation{Institute of Theoretical Physics, University of Warsaw, Pasteura 5, 02-093 Warsaw, Poland}
\affiliation{Centre for Quantum Technologies, National University of Singapore, 3 Science Drive 2, 117543 Singapore, Singapore}

\begin{abstract}
We investigate the Brout-Englert-Higgs mechanism of spontaneous symmetry breaking and show that, before symmetry breaking, the interaction of Higgs fields with massless gauge fields leads to the production of particles with negative squared mass.
\end{abstract}

\maketitle

%---------------------------------------------------
\section{Introduction}
%---------------------------------------------------

The celebrated Brout-Englert-Higgs mechanism~\cite{Anderson1963, Higgs1964, Englert1964, Guralnik1964} is crucial in explaining the non-zero mass of gauge bosons and fermions in the Standard Model. It involves a scalar field with a symmetric ``Mexican hat'' potential capable of spontaneous symmetry breaking~\cite{Srednicki2007}. The gauge 
bosons and fermions acquire masses through their interactions with the scalar field stabilized at a local minimum of the potential. Only after this stabilization takes place, the theory is quantized and excitations of the Higgs field around the minimum manifest themselves as Higgs particles.

In the usually adopted picture,~\cite{Larkoski2019}, the would-be tachyonic scalar field falls to a local minimum by itself due to the instability of the classical field configuration at the local maximum, where even a small disturbance will cause the field to fall in one direction, leading to symmetry breaking. Rolling down of the field is explained by means of long-wavelength modes of the field~\cite{Felder2001_1, Felder2001_2}, which are expected to grow exponentially in time. The short-wavelength modes are neglected in this description, as they do not exhibit such a growth.

Contrary to the long-wavelength modes of the field, the short-wavelength ones can be interpreted as particle excitations with real energies~\cite{Feinberg1967, Arons1968, Dhar1968, Paczos2023}. Physics of such particles was considered by Bilaniuk, Deshpande, and Sudarshan~\cite{Bilaniuk1962} and more recently in \cite{Dragan2020, Dragan2023}, while their covariant description in free quantum field theory was introduced in \cite{Paczos2023} by doubling the Hilbert space and thus overcoming the well-known mathematical obstacles~\cite{Arons1968, Dhar1968, Schwartz2018}.

In standard QFT, the action of the Lorentz group for standard quantum fields is represented in the Fock space ${\cal F} \equiv  \bigoplus_{n=0}^\infty S\left({\cal H}^{\otimes n}\right)$,	where ${\cal H}$ is a one-particle Hilbert space and $S$ is the symmetrization operator. In such a framework, the energy spectrum of negative-square-mass fields is unbounded from below, the vacuum state is frame-dependent and unstable, and the commutation rules are non-covariant. In \cite{Paczos2023} it was shown that these issues are a result of misrepresenting the Lorentz group in a too small Hilbert space. By doubling this space to ${\cal F}\otimes{\cal F}^\star$, it is possible to establish an explicitly covariant framework that allows for the proper quantization of the tachyonic fields, eliminating all of these issues. The field operator $\hat{\Phi}$ acting in the twin space ${\cal F}\otimes{\cal F}^\star$ was shown to be a scalar under the representation $U(\Lambda)$ of the Lorentz group: $\boldsymbol{\hat{\Phi}}(\Lambda^{-1} x) = U(\Lambda)^{-1} \boldsymbol{\hat{\Phi}}(x) U(\Lambda)$. The authors of \cite{Paczos2023} showed that the twin-space formalism, similar to the two-state vector formalism developed by Aharonov {\it et al.} \cite{Aharonov1964}, is needed for covariant transformation of probability amplitudes, but reduces to the standard QFT formalism in a fixed reference frame.

Based on these results, in this work, we consider the quantum theory of Higgs-like fields interacting with massless fields and describe the possible short-wavelength particle production processes that occur during the initial symmetric phase just before the symmetry is spontaneously broken. We investigate a simple model involving a negative square mass field and a massless scalar field interacting via a scalar Yukawa potential. This model is a natural proxy for conceptually similar models of the interaction between the Higgs field of the Standard Model and the massless gauge or matter fields before the electroweak symmetry breaking. We show that such coupling leads to the spontaneous emission of short-wavelength particles by massless particles, which may be viewed as a sign of the destabilization of the quantum tachyonic vacuum in the presence of interactions. In contrast to the common finite-temperature initial-state assumption for Higgs-like fields~\cite{Dolan1974, Weinberg1974}, we consider the simpler case of a vacuum (zero-temperature) initial state. We further restrict attention to early-time dynamics, neglecting the backreaction from exponentially growing long-wavelength modes mediated by the $\lambda\varphi^4$ interaction. Within this controlled setting, we can transparently isolate the short-wavelength particle-production mechanism.

%---------------------------------------------------
\section{Symmetry breaking model}
%---------------------------------------------------
Let us consider the simplest model of spontaneous symmetry breaking with the Lagrangian
\begin{equation}\label{Lagrangian}
    \mathcal{L}=-\frac{1}{2}\partial^\mu\varphi\partial_\mu\varphi+\frac{1}{2}m^2\varphi^2-\frac{1}{24}\lambda\varphi^4
\end{equation}
describing the dynamics of a real scalar field $\varphi$. The unusual feature of this Lagrangian, distinguishing it from the Lagrangian of the ordinary scalar field, is the plus sign in front of the mass term. This makes the scalar potential $V(\varphi)=-m^2\varphi^2/2+\lambda\varphi^4/24$ have a local maximum at $\varphi=0$, and two local minima at $\varphi=\pm\sqrt{6m^2/\lambda}$. Note that this potential is symmetric under $\varphi\to-\varphi$. Typically, one expands the field around one of the local minima, after which the symmetry is lost --- this is referred to as the spontaneous symmetry breaking. 

Notably, the field around the point $\varphi=0$, for small values of the field $\varphi$, is formally tachyonic. The simplest way to see this is to neglect the term $\propto\lambda\varphi^4$ in the Lagrangian~\eqref{Lagrangian} and write the corresponding field equation
\begin{equation}\label{field equation}
    \left(\partial^\mu\partial_\mu-m^2\right)\varphi=0.
\end{equation}
Making use of the correspondence between the energy and momentum in classical and quantum theory, i.e., $E\leftrightarrow \mathrm{i}\partial_t$ and $\boldsymbol{p}\leftrightarrow-\mathrm{i}\nabla$, we conclude that Eq.~\eqref{field equation} corresponds to the energy-momentum relation
\begin{equation}
    \label{energymomentum}
    E^2=\boldsymbol{p}^2-m^2.
\end{equation}
Formula~\eqref{energymomentum} gives rise to imaginary energies for $|\boldsymbol{p}|<m$ (long-wavelength modes), and real energies for $|\boldsymbol{p}|\geq m$ (short-wavelength modes). Only the latter can be interpreted in terms of particles~\cite{Bilaniuk1962, Feinberg1967, Arons1968, Dhar1968, Dragan2020, Dragan2023, Paczos2023}.

Long-wavelength modes (with $|\boldsymbol{p}|<m$) are expected to grow exponentially in time (due to their imaginary energies) during the symmetry-breaking process, and to dominate the behavior of the field~\cite{Felder2001_1, Felder2001_2}. Therefore, the short-wavelength modes, despite their interesting particle interpretation, are completely ignored. Here, we focus on the short-wavelength modes instead and study their interaction with other fields. We assume that the particles interacting with the scalar field gain mass only due to the Brout-Englert-Higgs mechanism; therefore, around $\varphi=0$, they are massless. The interaction between massless particles and tachyons enables processes that would be forbidden if we considered only positive square mass particles. An example of such a process is the spontaneous emission of a tachyon by a massless particle. We will analyze this process as a possible origin of the spontaneous symmetry breaking. As a toy model, we consider a Yukawa-type interaction with the interaction Hamiltonian
\begin{equation}\label{Yukawa interaction}
    \mathcal{H}_\text{int}=g\varphi\psi^2
\end{equation}
where $\psi$ is a massless real scalar field.

The quantization of the tachyonic field $\varphi$ restricted to short-wavelength modes has been performed in~\cite{Paczos2023}, where it was shown that to preserve Lorentz covariance of the theory, one has to describe the quantum states of the tachyonic field in an extended Fock space 
with a double set of states that can be interpreted as in and out states.
However, in a fixed reference frame, one can separate the two Fock spaces 
and perform calculations in complete analogy to the standard quantum field theory. Then, one can use the effective field operator
\begin{align}
\label{singlefockoperator}
    \hat{\varphi}(t,\boldsymbol{r}) = \int_{|\boldsymbol{k}|\geq m}^{} \text{d}^3\boldsymbol{k}\, \left(
u_{\boldsymbol{k}}(t,\boldsymbol{r})\,\hat{a}_{\boldsymbol{k}}  + u^*_{\boldsymbol{k}}(t,\boldsymbol{r})\,\hat{a}^{\dagger}_{\boldsymbol{k}}   \right),
\end{align}
where $u_{\boldsymbol{k}}(t,\boldsymbol{r})$ are the mode solutions to the tachyonic Klein-Gordon equation~\eqref{field equation}, orthonormal with respect to the inner product $(\chi,\eta)=i\int_{t=0}\text{d}^3 \boldsymbol{r} \,\chi^*\overleftrightarrow{\partial_t}\eta$, and $\hat{a}_{\boldsymbol{k}}$ and $\hat{a}^{\dagger}_{\boldsymbol{k}}$ are the tachyonic ladder operators satisfying standard commutation relations
\begin{equation}
\label{eqn:canonical_commutation_relations}
[\hat{a}_{\boldsymbol{k}}, \hat{a}^{\dagger}_{\boldsymbol{l}}] = 2 \omega_k(2\pi)^3\delta^{(3)}(\boldsymbol{k}-\boldsymbol{l}).
\end{equation}
The mode solutions are given by
\begin{equation}
\label{normalmodes}
u_{\boldsymbol{k}}(t, \boldsymbol{r}) = \frac{1}{(2\pi)^3 2\Omega_k} e^{i(\boldsymbol{k}\cdot\boldsymbol{r}-\Omega_k t)},
\end{equation}
where $\Omega_{\boldsymbol{k}}$ satisfies the tachyonic dispersion relation $\Omega_{\boldsymbol{k}}^2 = \boldsymbol{k}^2-m^2$ with $|\boldsymbol{k}|\geq m$.

On the other hand, the massless field $\psi$ is quantized in the standard way, with the corresponding field operator
\begin{equation}
    \hat{\psi}(x)=\int\mathrm{d}^3\boldsymbol{k}\left(v_{\boldsymbol{k}}(x)\hat{b}_{\boldsymbol{k}}+v_{\boldsymbol{k}}^*(x)\hat{b}^\dagger_{\boldsymbol{k}}\right).
\end{equation}
Here $v_{\boldsymbol{k}}$ are mode solutions to the massless field equation $\partial^\mu\partial_\mu\psi=0$, namely
\begin{equation}\label{v mode definition}
    v_{\boldsymbol{k}}(x)=\frac{1}{(2\pi)^3 2|\boldsymbol{k}|}\mathrm{e}^{-i(|\boldsymbol{k}|t-\boldsymbol{k}\cdot\boldsymbol{x})},
\end{equation}
and $\hat{b}_{\boldsymbol{k}}$, $\hat{b}_{\boldsymbol{k}}^\dagger$ are the ladder operators satisfying the standard commutation rules
\begin{equation}\label{eqn:canonical_commutation_relations_2}
    [\hat{b}_{\boldsymbol{k}},\hat{b}^\dagger_{\boldsymbol{l}}]=2|\boldsymbol{k}|(2\pi)^3 \delta^{(3)}(\boldsymbol{k}-\boldsymbol{l})
\end{equation}
with all the other commutators vanishing.

%---------
\section{Particle emission process}
%---------

We focus on tachyon emission in the first-order perturbative expansion, where a massless particle of momentum $\boldsymbol{k}$ changes its momentum into $\boldsymbol{p}$ while emitting a tachyon of momentum $\boldsymbol{l}$. This corresponds to the initial state $\ket{i}\propto\hat{b}^\dagger_{\boldsymbol{k}}\ket{0}$ and the final state $\ket{f}\propto\hat{a}^\dagger_{\boldsymbol{l}}\hat{b}^\dagger_{\boldsymbol{p}}\ket{0}$. The $S$-matrix element for this process is given by
\begin{equation}
    S_{fi}=-i\bra{f}\int\mathrm{d}^4x\mathcal{H}_\text{int}(x)\ket{i}.
\end{equation}
After inserting the exact form of the interaction Hamiltonian from Eq.~\eqref{Yukawa interaction} and using the commutation relations~\eqref{eqn:canonical_commutation_relations} and~\eqref{eqn:canonical_commutation_relations_2}, we obtain (see Appendix~\ref{app: decay rate} for details)
\begin{equation}
    S_{fi}=-ig(2\pi)^4\delta^{(4)}(p+l-k).
\end{equation}
Let us point out that the resulting scattering amplitude $\mathcal{A}_{fi}\equiv g$ is Lorentz invariant, as required.

Given the scattering amplitude we can calculate the rate at which tachyons are emitted --- the decay width:
\begin{equation}
    \begin{split}
        \Gamma\equiv\frac{1}{2 |\boldsymbol{k}|} \int_{|\boldsymbol{l}|\geq m} &\frac{\mathrm{d}^3 \boldsymbol{l}}{(2 \pi)^3 2 \Omega_{\boldsymbol{l}}}\frac{\mathrm{d}^3 \boldsymbol{p}}{(2 \pi)^3 2 |\boldsymbol{p}|}|\mathcal{A}_{fi}|^2\times\\
        &\times(2 \pi)^4 \delta^{(4)}\left(k-l-p\right).
    \end{split}
\end{equation}
We have formally included here the constraint $|\boldsymbol{l}|\geq m$ restricting the tachyonic momenta [compare with Eq.~\eqref{singlefockoperator}]. However, it can be verified (see Appendix~\ref{app: decay rate}) that it is automatically satisfied due to the support of the Dirac delta function.

Performing the integral over the tachyonic momentum~$\boldsymbol{l}$ we obtain
\begin{equation}
    \label{rate_intermediate_1}
    \Gamma=\frac{g^2}{2 |\boldsymbol{k}|} \int \frac{\mathrm{d}^3 \boldsymbol{p}}{(2 \pi)^3}\frac{1}{2|\boldsymbol{p}|}\frac{1}{2 \Omega_{\boldsymbol{k}-\boldsymbol{p}}}2 \pi \delta(|\boldsymbol{k}|-|\boldsymbol{p}|-\Omega_{\boldsymbol{k}-\boldsymbol{p}})\, .
\end{equation}
In the reference frame where $\boldsymbol{k} = (0,0,\abs{\boldsymbol{k}})$, we parametrize $\boldsymbol{p}$ using spherical coordinates defined by angles $\theta$ (polar) and $\varphi$ (azimuthal), which determine the direction of motion of the final massless particle. Since the emitted tachyon has a positive energy, the energy $|\boldsymbol{p}|$ of the massless particle in the final state must be smaller than the initial energy, $|\boldsymbol{p}|<|\boldsymbol{k}|$. This implies that $\theta$ is larger than the boundary angle $\theta_0$ given by
\begin{equation}\label{boundary angle 0}
    \cos\theta_0=1-\frac{m^2}{2|\boldsymbol{k}|^2} \, .
\end{equation}
This results in the following restriction in the decay width integral \eqref{rate_intermediate_1}:
\begin{equation}
    \label{gamma_last_integral}
    \Gamma = \frac{g^2}{16\pi \abs{\boldsymbol{k}}} \frac{m^2}{2 \abs{\boldsymbol{k}}^2}  \int_{-1}^{1 - \frac{m^2}{2 \abs{\boldsymbol{k}}^2}}\frac{\mathrm{d}(\cos \theta)}{(1- \cos \theta)^2}.
\end{equation}
Performing the integral we obtain
\begin{equation}\label{rate of emission}
    \Gamma = \begin{cases}
        \frac{g^2}{16\pi \abs{\boldsymbol{k}}} \left( 1 - \frac{m^2}{4 \abs{\boldsymbol{k}}^2} \right), & \text{if $|\boldsymbol{k}|>m/2$},\\
        0, & \text{otherwise}.
    \end{cases}
\end{equation}
Therefore, the emission of a tachyon with mass $m$ is only possible if the initial massless particle has energy greater than $m/2$. This condition is clearly frame-dependent since a boosted reference frame can always be found in which it is no longer satisfied, causing the decay width to vanish.

Let us elaborate on this interesting observation. Unlike in standard QFT,
in which Lorentz covariance requires that the decay width is inversely proportional to the energy of the decaying particle, $\Gamma\propto |\boldsymbol{k}|^{-1}$ (see e.g.~\cite{Modanese1995}), our result \eqref{rate of emission} contains a term that scales as $|\boldsymbol{k}|^{-3}$.
This originates from the factor:
\begin{equation}
    \label{noncov factor}
    I=\frac{m^2}{2|\boldsymbol{k}|^2}\int_{-1}^{\cos\theta_0}\frac{\mathrm{d}(\cos\theta)}{(1-\cos\theta)^2}=1-\frac{m^2}{4|\boldsymbol{k}|^2}.
\end{equation}
The breaking of covariance arises from the requirement that the emitted tachyon has positive energy. As previously argued in \cite{Bilaniuk1962, Feinberg1967, Paczos2023}, an emission process in one reference frame may appear as an absorption process in another. Since changing the energy $|\boldsymbol{k}|$ of the initial particle corresponds to applying a Lorentz boost along the $z$-direction, certain events will enter or leave the phase space integral as $|\boldsymbol{k}|$ changes.

To illustrate this issue, consider a reference frame boosted by $\beta$ along the positive $z$-direction relative to the initial inertial frame where the massless particle has energy $\abs{\boldsymbol{k}}$. In the boosted reference frame, the initial energy $|\boldsymbol{k}'|$ is:
\begin{equation}
    \label{boost k}
    |\boldsymbol{k}'|=|\boldsymbol{k}|\sqrt{\frac{1-\beta}{1+\beta}}.
\end{equation}
Let us denote by $\tilde{\theta}_0$ the boundary angle in the boosted reference frame (analogous to $\theta_0$ in the original frame), given by
\begin{equation}
    \label{angle in new frame}
    \cos\tilde{\theta}_0=1-\frac{m^2}{2|\boldsymbol{k}'|^2}=1-\frac{m^2}{2|\boldsymbol{k}|^2}\frac{1+\beta}{1-\beta}.
\end{equation}
Let us also introduce the final momentum of the massless particle $\boldsymbol{p}_0$ corresponding to the boundary angle $\theta_0$ in the original frame. Denote by $\boldsymbol{p}_0'$ the boosted counterpart of $\boldsymbol{p}_0$ and by $p_{0z}'$ its $z$-component, and define angle $\theta_0'$ as follows:
\begin{equation}
\label{boosted boundary angle 0}
    \cos\theta_0'\equiv\frac{p_{0z}'}{|\boldsymbol{p}_0'|}=\frac{\cos\theta_0-\beta}{1-\beta\cos\theta_0}.
\end{equation}
Inserting $\cos\theta_0$ from Eq.~\eqref{boundary angle 0} to Eq.~\eqref{boosted boundary angle 0} reveals that $\cos\theta_0'\neq\cos\tilde{\theta}_0$. In particular, for $\beta>0$ we have $\cos\theta_0'>\cos\tilde{\theta}_0$, indicating that some of the emission events from the original reference frame are not accounted for in the new frame as they become classified as absorption events.

This discrepancy leads to the non-covariance of the decay widths. Indeed, we define the decay width $\tilde{\Gamma}$ in the boosted reference frame as
\begin{equation}\label{boosted gamma}
\begin{split}
    &\tilde{\Gamma}\equiv\frac{g^2}{16\pi|\boldsymbol{k}'|}\tilde{I},\\
    &\tilde{I}\equiv\frac{m^2}{2|\boldsymbol{k}'|^2}\int_{-1}^{\cos\tilde{\theta}_0}\frac{\mathrm{d}(\cos\theta)}{(1-\cos\theta)^2}=1-\frac{m^2}{4|\boldsymbol{k}'|^2},
\end{split}
\end{equation}
using angle $\tilde{\theta}_0$ to ensure that only proper emission processes are accounted for. The ratio $\tilde{\Gamma}/\Gamma$ is then given by
\begin{equation}\label{gamma ratio}
    \frac{\tilde{\Gamma}}{\Gamma}=\frac{|\boldsymbol{k}|}{|\boldsymbol{k}'|}\frac{1-m^2/4|\boldsymbol{k}|^2}{1-m^2/4|\boldsymbol{k}'|^2}\neq\frac{|\boldsymbol{k}|}{|\boldsymbol{k}'|},
\end{equation}
highlighting the non-covariance mentioned above.

Now define the boosted-frame ``interaction rate'' $\Gamma'$ using $\theta_0'$ instead of $\tilde{\theta}_0$:
\begin{equation}\label{fake noncov term}
\begin{split}
    &\Gamma'\equiv\frac{g^2}{16\pi|\boldsymbol{k}'|}I',\\
    &I'\equiv\frac{m^2}{2|\boldsymbol{k}'|^2}\int_{-1}^{\cos\theta_0'}\frac{\mathrm{d}(\cos\theta)}{(1-\cos\theta)^2}=1-\frac{m^2}{4|\boldsymbol{k}|^2}=I.
\end{split}
\end{equation}
This definition ensures that in both reference frames, we take into account the same set of events (even though some of them are absorption events in the boosted frame). With this definition, the ratio $\Gamma'/\Gamma$ is equal to
\begin{equation}\label{fake gamma ratio}
    \frac{\Gamma'}{\Gamma}=\frac{|\boldsymbol{k}|}{|\boldsymbol{k}'|},
\end{equation}
as would be expected from Lorentz covariant decay widths. This shows that the non-covariance of the decay width~\eqref{rate of emission} is not a problem of the formalism that we use, but stems from the frame-dependent notion of tachyon emission.

%-------------------
\section{Discussion}
%-------------------

We have investigated the interaction between the short-wavelength modes of a Higgs-like field in the unbroken-symmetry phase with other fields gaining mass due to the Brout-Englert-Higgs mechanism. The interaction leads to spontaneous excitation of the tachyonic field modes by the excitations of the other fields. We have derived the decay widths of the massless fields and showed that their non-covariance is not a flaw of the formalism, but rather a fundamental consequence of the nature of tachyonic interactions. Specifically, the notion of decay itself becomes a frame-dependent concept. As we have argued in \cite{Paczos2023}, the observed non-covariance of certain observables arises from the fact that the Lorentz group does not preserve the standard Fock space $\mathcal{F}$ in the tachyonic case. Instead, Lorentz invariance is only restored when considering action on the full twin space $\mathcal{F} \otimes \mathcal{F}^\star$, where in-states and out-states are connected through Lorentz boosts. Consequently, a process identified as the emission of a tachyon in one reference frame may appear as absorption in another, since we use the states from Fock subspaces $\mathcal{F}$ to define what is the cause of absorption or emission in a given interaction. This feature is unique to tachyonic quantum fields; in contrast, for conventional quantum fields, the Lorentz group preserves each Fock subspace individually, precluding such transformations.

Furthermore, while the fundamental laws of physics and the resulting probability amplitudes from $S$-matrix formalism remain Lorentz invariant~\cite{Paczos2023}, the physical systems we observe often establish preferred reference frames, breaking Lorentz symmetry in practice. One prominent example is the cosmic microwave background (CMB) radiation, which identifies frames in which it appears isotropic. Thermal radiation, unlike its zero-temperature limit (the zero-point spectrum, with an energy density scaling as $\mathrm{d}u \propto \omega^3 \mathrm{d}\omega$), is not Lorentz invariant. This suggests that processes like those discussed here, particularly the emission of tachyons, could contribute to violations of Lorentz symmetry in the early universe. Such processes might leave observable imprints in the final state of radiation, manifesting as non-trivial contributions to the power spectra, irrespective of subsequent decay or interaction channels.

Our calculation serves as a means to isolate the leading-order phenomenon, while the higher-order calculations involving tachyonic fields, are yet to be fully understood, which was recently pointed out in an interesting work by Jodłowski \cite{Jodlowski:2024rut}. Although our analysis focuses on a simplified model of the initial phase of spontaneous symmetry breaking, the results are expected to generalize to more realistic scenarios. For instance, a complex tachyonic field interacting through a Yukawa interaction with a massless Dirac field would serve as an analog to the Higgs mechanism for fermions~\cite{Larkoski2019}. Thus, the spontaneous emission of tachyons by massless particles, as analyzed here, may serve as a fundamental mechanism or instability-driving spontaneous symmetry breaking in the Higgs mechanism. Exploring the broader implications of this phenomenon and extending it to other physical models will be the focus of future work.

\section*{Acknowledgements}
J.P. is grateful to Fawad Hassan and Magdalena Zych for useful discussions. J.P. acknowledges the Knut and Alice Wallenberg Foundation through a Wallenberg Academy Fellowship No. 2021.0119.

\bibliographystyle{apsrev4-2}
\bibliography{ref}

\onecolumngrid

\appendix

%---------------------------------------------------
\section{Decay width}\label{app: decay rate}
%---------------------------------------------------
We analyze the emission process $\psi\to\psi\varphi$, where a massless particle emits a tachyon. The corresponding (appropriately normalized) initial and final states are given by
\begin{equation}     
    \ket{i}=\sqrt{(2\pi)^3 2|\boldsymbol{k}|}\hat{b}^\dagger_{\boldsymbol{k}}\ket{0},\qquad \ket{f}=\sqrt{(2\pi)^3 2\Omega_{\boldsymbol{l}}}\hat{a}^\dagger_{\boldsymbol{l}}\sqrt{(2\pi)^3 2|\boldsymbol{p}|}\hat{b}^\dagger_{\boldsymbol{p}}\ket{0}.
\end{equation}
Here $\boldsymbol{k}$, $\boldsymbol{p}$, and $\boldsymbol{l}$ denote the momenta of massless particles in the initial and final states, and the tachyonic momentum, respectively. In the first-order perturbation theory, the contribution to the absorption process is given by
\begin{equation}
\begin{split}
    S_{fi}\approx&-i\bra{f}\int\mathrm{d}^4x\mathcal{H}_\text{int}(x)\ket{i}=-ig\bra{f}\int\mathrm{d}^4x\hat{\varphi}(x)\left(\hat{\psi}(x)\right)^2\ket{i}\\
    =&-ig\sqrt{(2\pi)^9 8\Omega_{\boldsymbol{l}}|\boldsymbol{k}||\boldsymbol{p}|}\bra{0}\hat{b}_{\boldsymbol{p}}\hat{a}_{\boldsymbol{l}}\int\mathrm{d}^4x\hat{\varphi}(x)\left(\hat{\psi}(x)\right)^2\hat{b}^\dagger_{\boldsymbol{k}}\ket{0}.
\end{split}
\end{equation}
Now, we use the fact that
\begin{equation}
    [\hat{b}_{\boldsymbol{p}},\hat{\psi}(x)]=v^*_{\boldsymbol{p}}(x),\qquad [\hat{\psi}(x),\hat{b}^\dagger_{\boldsymbol{k}}]=v_{\boldsymbol{k}}(x),\qquad [\hat{a}_{\boldsymbol{l}},\hat{\varphi}(x)]=u^*_{\boldsymbol{l}}(x),
\end{equation}
to get
\begin{equation}\label{amplitude}
    S_{fi}\approx-ig\sqrt{(2\pi)^9 8\Omega_{\boldsymbol{l}}|\boldsymbol{k}||\boldsymbol{p}|}\int\mathrm{d}^4xv^*_{\boldsymbol{p}}(x)u^*_{\boldsymbol{l}}(x)v_{\boldsymbol{k}}(x)=-ig(2\pi)^4\delta^{(4)}(p+l-k)\equiv(2\pi)^4\delta^{(4)}(k+l-p)(-i\mathcal{A}_{fi}),
\end{equation}
where $\mathcal{A}_{fi}\equiv g$ is the amplitude of the process.

According to the general prescription, the decay width of the particle of four-momentum $k$ into $n$ particles indexed by $j$ is given by
\begin{equation}\label{decay rate}
    \Gamma=\frac{1}{2 E_{\boldsymbol{k}}} \int \prod_{j=1}^{n}\left(\frac{\mathrm{d}^3 \boldsymbol{p}_j}{(2 \pi)^3 2 E_{\boldsymbol{p}_j}}\right)\left|\mathcal{A}_{f i}\right|^2(2 \pi)^4 \delta^{(4)}\left(k-\sum_j p_j\right) ,
\end{equation}
where $E_{\boldsymbol{q}}$ is the energy of a particle with four-momentum $q$. In our case the formula~\eqref{decay rate} can be rewritten as
\begin{equation}\label{rate}
    \Gamma=\frac{1}{2 |\boldsymbol{k}|} \int_{|\boldsymbol{l}|\geq m} \frac{\mathrm{d}^3 \boldsymbol{l}}{(2 \pi)^3 2 \Omega_{\boldsymbol{l}}}\frac{\mathrm{d}^3 \boldsymbol{p}}{(2 \pi)^3 2 |\boldsymbol{p}|} g^2(2 \pi)^4 \delta^{(3)}\left(\boldsymbol{k}-\boldsymbol{l}-\boldsymbol{p}\right)\delta(|\boldsymbol{k}|-|\boldsymbol{p}|-\Omega_{\boldsymbol{l}}).
\end{equation}
Let us choose the reference frame in which
\begin{equation}
    \boldsymbol{k}=(0,0,|\boldsymbol{k}|),
\end{equation}
i.e., the initial massless particle moves along the $z$-direction. Let us also introduce the angles $\theta$ and $\varphi$ that determine the direction of motion of the final massless particle, so that
\begin{equation}
    \boldsymbol{p}=(|\boldsymbol{p}|\sin\theta\cos\varphi,|\boldsymbol{p}|\sin\theta\sin\varphi,|\boldsymbol{p}|\cos\theta).
\end{equation}
We can now integrate \eqref{rate} over $d^3 \boldsymbol{l}$ using the delta function $\delta^{(3)}\left(\boldsymbol{k}-\boldsymbol{l}-\boldsymbol{p}\right)$, which will fix $\boldsymbol{l}=\boldsymbol{k}-\boldsymbol{p}$. Hence, we have
\begin{equation}
    \label{rate2}
    \Gamma=\frac{g^2}{2 |\boldsymbol{k}|} \int_{|\boldsymbol{k}-\boldsymbol{p}|\geq m} \frac{\mathrm{d}^3 \boldsymbol{p}}{(2 \pi)^3}\frac{1}{2|\boldsymbol{p}|}\frac{1}{2 \Omega_{\boldsymbol{k}-\boldsymbol{p}}}2 \pi \delta(|\boldsymbol{k}|-|\boldsymbol{p}|-\Omega_{\boldsymbol{k}-\boldsymbol{p}}),
\end{equation}
where $\mathrm{d}^3 \boldsymbol{p}=|\boldsymbol{p}|^2\mathrm{d}|\boldsymbol{p}|\sin\theta \mathrm{d}\theta \mathrm{d}\varphi$. The integration over $\mathrm{d}\varphi$ can be done trivially since the integrand does not depend on $\varphi$ --- this contributes a factor of $2\pi$. The integral over $\mathrm{d}|\boldsymbol{p}|$ is more elaborate and requires working out the explicit form of the delta function argument. The delta function $\delta(|\boldsymbol{k}|-|\boldsymbol{p}|-\Omega_{\boldsymbol{k}-\boldsymbol{p}})$ is non-zero only if the following condition is satisfied:
\begin{equation}\label{condition}
    |\boldsymbol{k}|-|\boldsymbol{p}|-\sqrt{|\boldsymbol{k}-\boldsymbol{p}|^2-m^2}=0.
\end{equation}
We have
\begin{equation}
    \boldsymbol{k}-\boldsymbol{p}=(-|\boldsymbol{p}|\sin\theta\cos\varphi,-|\boldsymbol{p}|\sin\theta\sin\varphi,|\boldsymbol{k}|-|\boldsymbol{p}|\cos\theta),
\end{equation}
which means
\begin{equation}\label{absolute value}
    |\boldsymbol{k}-\boldsymbol{p}|^2=|\boldsymbol{k}|^2+|\boldsymbol{p}|^2-2|\boldsymbol{k}||\boldsymbol{p}|\cos\theta.
\end{equation}
Thus, Eq.~\eqref{condition} is equivalent to
\begin{equation}
    |\boldsymbol{k}|-|\boldsymbol{p}|=\sqrt{|\boldsymbol{k}|^2+|\boldsymbol{p}|^2-2|\boldsymbol{k}||\boldsymbol{p}|\cos\theta-m^2}.
\end{equation}
We square both sides to get
\begin{equation}\label{delta condition}
    |\boldsymbol{k}|^2+|\boldsymbol{p}|^2-2|\boldsymbol{k}||\boldsymbol{p}|=|\boldsymbol{k}|^2+|\boldsymbol{p}|^2-2|\boldsymbol{k}||\boldsymbol{p}|\cos\theta-m^2,
\end{equation}
which implies
\begin{equation}
    |\boldsymbol{p}|=\frac{m^2}{2|\boldsymbol{k}|(1-\cos\theta)}\leq |\boldsymbol{k}|.
\end{equation}
The last condition $|\boldsymbol{p}|\leq |\boldsymbol{k}|$ must be satisfied since we are considering the actual emission of the tachyon --- the process in which the massless particle can only lose energy, which is also required by Eq. \eqref{condition}. This restricts the angle $\theta$ as follows:
\begin{equation}\label{angle boundary}
    \cos\theta\leq1-\frac{m^2}{2|\boldsymbol{k}|^2}.
\end{equation}
Making use of the delta function properties, we write
\begin{equation}
\begin{split}
    \delta(|\boldsymbol{k}|-|\boldsymbol{p}|-\Omega_{\boldsymbol{k}-\boldsymbol{p}})=&\frac{\delta\left(|\boldsymbol{p}|-\frac{m^2}{2|\boldsymbol{k}|(1-\cos\theta)}\right)}{\left|1+\frac{|\boldsymbol{p}|-|\boldsymbol{k}|\cos\theta}{\sqrt{|\boldsymbol{k}|^2+|\boldsymbol{p}|^2-2|\boldsymbol{k}||\boldsymbol{p}|\cos\theta-m^2}}\right|}
    =\frac{\delta\left(|\boldsymbol{p}|-\frac{m^2}{2|\boldsymbol{k}|(1-\cos\theta)}\right)}{\left|1+\frac{|\boldsymbol{p}|-|\boldsymbol{k}|\cos\theta}{|\boldsymbol{k}|-|\boldsymbol{p}|}\right|}\\=&\delta\left(|\boldsymbol{p}|-\frac{m^2}{2|\boldsymbol{k}|(1-\cos\theta)}\right)\frac{|\boldsymbol{k}|-|\boldsymbol{p}|}{|\boldsymbol{k}|(1-\cos\theta)}.
\end{split}
\end{equation}
Let us also notice that~\eqref{absolute value} combined with~\eqref{delta condition} imply
\begin{equation}
    |\boldsymbol{k}-\boldsymbol{p}|=\sqrt{m^2+(|\boldsymbol{k}|-|\boldsymbol{p}|)^2}\geq m.
\end{equation}
With this in hand, we can perform the integrals over $\mathrm{d}|\boldsymbol{p}|$ and $\mathrm{d}\theta$ in \eqref{rate2} to get
\begin{equation}\label{decay rate 2}
    \begin{split}
        \Gamma =& \frac{g^2}{16\pi \abs{\boldsymbol{k}}} \int_{\abs{\boldsymbol{k} - \boldsymbol{p}} \geq m} \mathrm{d}\abs{\boldsymbol{p}} \mathrm{d}\theta \frac{\abs{\boldsymbol{p}} \sin \theta}{\abs{\boldsymbol{k}} (1- \cos \theta)}\delta\left( \abs{\boldsymbol{p}} - \frac{m^2}{2\abs{\boldsymbol{k}
        } (1- \cos \theta)} \right)\Theta\left(1-\frac{m^2}{2|\boldsymbol{k}|^2}-\cos\theta\right) \\  
        =& \frac{g^2}{16\pi \abs{\boldsymbol{k}}} \frac{m^2}{2 \abs{\boldsymbol{k}}^2} \int_{-1}^{1 - \frac{m^2}{2 \abs{\boldsymbol{k}}^2}} \frac{\mathrm{d}(\cos \theta)}{(1- \cos \theta)^2}
        = \frac{g^2}{16\pi \abs{\boldsymbol{k}}} \left( 1 - \frac{m^2}{4 \abs{\boldsymbol{k}}^2} \right) \Theta\left(2-\frac{m}{|\boldsymbol{k}|}\right)
    \end{split}
\end{equation}
where $\Theta(\cdot)$ is the Heaviside step function and appears due to inequalities $-1 \leq 1 - m^2 / (2 \abs{\boldsymbol{k}}^2 ) \leq 1$. This decay rate is non-zero for $m/|\boldsymbol{k}|<2$.

% \bibliographystyle{elsarticle-harv} 
% \bibliography{ref}



%% For example:

%% \bibitem[Aladro et al.(2015)]{Aladro15} Aladro, R., Martín, S., Riquelme, D., et al. 2015, \aas, 579, A101

%%\end{thebibliography}

\end{document}